\documentclass[manuscript]{raa}
\usepackage{graphicx,times}
\usepackage{natbib}
\usepackage{longtable}
\usepackage{amsmath}
\usepackage{amssymb}

\providecommand{\bm}[1]{\mbox{\boldmath$#1$\unboldmath}}

\begin{document}

   \title{Sequential clustering of star formation in IC 1396
}

 \volnopage{ {\bf 20xx} Vol.\ {\bf 9} No. {\bf XX}, 000--000}
   \setcounter{page}{1}

   \author{Ya Fang Huang
      \inst{}
   \and Jin Zeng Li
      \inst{} }

   \institute{National Astronomical Observatories, Chinese Academy of Sciences,
             Beijing 100012, China; {\it huangyf@nao.cas.cn}\\
\vs \no
   {\small Received [year] [month] [day]; accepted [year] [month] [day] }
}

\abstract{We present in this paper a comprehensive study of the H~\textsc{ii} region IC~1396 and its star formation activity,
in which multi-wavelength data ranging from the optical to the near- and far-infrared were employed. The surface density distribution of all the 2MASS sources with certain detection toward IC 1396 indicates the existence
of a compact cluster spatially consistent with the position of the exciting source of the H~\textsc{ii} region, HD~206267. The spatial distribution of the infrared excessive emission sources selected based on archived 2MASS data reveals the existence of four sub-clusters in this region. One is in association with the open cluster Trumpler~37. The other three
are found to be spatially coincident with the bright rims of the H~\textsc{ii} region.
All the excessive emission sources in the near infrared are cross-identified with the AKARI IRC data, an analysis of the spectral energy distributions (SEDs) of
the resultant sample leads to the identification of 8 CLASS I, 15 CLASS II and 15 CLASS III sources in IC~1396. Optical identification
of the sample sources with R magnitudes brighter than 17 mag corroborates the results from the SED analysis. Based on the spatial distribution of
the infrared young stellar objects at different evolutionary stages, the surrounding sub-clusters located in the bright rims are
believed to be younger than the central one. This is consistent with a scenario of sequential star formation in this region.
Imaging data of a dark patch in IC~1396 by Herschel SPIRE, on the other hand, indicate the presence of two far-infrared cores in LDN 1111,
which are likely new generation protostellar objects in formation. So we infer that the star formation process in this H~\textsc{ii}
region was not continuous but episodic.
}

   \authorrunning{Huang \& Li}
   \titlerunning{Sequential clustering of star formation in IC 1396}  
   \maketitle


%
%
\section{Introduction}           
\label{sect:intro}

Molecular clouds \citep{2003ARA&A..41...57L} are essential sites of clustered star
formation and the formation of massive stars. Cepheus OB2 (Cep OB2), known to be located in the local
(Orion) spiral arm, is divided into two subgroups
\citep{1968ApJ...154..923S}: 1) Cep~OB2a contains 75 O- and B-stars that spread over a
large area, between $100^\circ<l<106^\circ$ and $+2^\circ<b<+8^\circ$. It has an estimated
age of 6-7 million years; 2) Cep OB2b, namely Trumpler 37 (Tr~37), has an age of 3 million
years and is among the youngest known open clusters \citep{2008hsf1.book..136K}.

Tr~37 is associated with the shell-like H~\textsc{ii} region,
IC~1396. IC~1396 is located at the southwest tip of Cep~OB2,
just above the galactic plane at $l=99.3^\circ$ and $b=3.74^\circ$.
This H~\textsc{ii} region is mainly excited by the central O6 star
HD~206267. As an extended H~\textsc{ii} region ($\sim3^\circ$) at about 800~pc
\citep{1968ApJ...154..923S}, it harbors a significant population of
Pre-Main Sequence stars (PMS) with low- to intermediate masses.

Large-scale observations of IC~1396 with rotational CO lines
\citep{1995ApJ...447..721P,1996A&A...309..581W} and the radio map
\citep{1980A&A....88..285M,1980A&A....89..180W} have uncovered a
ring-like and scattered structure. The appearance of the
H~\textsc{ii} region is dominated by a high degree of fragmentation
into many dark and bright-rimmed globules. Near-infrared (IR)
observations of IC~1396 and extinction maps obtained from Two Micron
All Sky Survey (2MASS) data reveal star-formation activity and large
numbers of globules in this region \citep{2005A&A...432..575F}. Some
of the densest regions in IC~1396 have already been investigated in
detail (IC~1396N: \cite{2007A&A...468L..33N,2010ApJ...717.1067C};
\cite{2007ApJ...654..316G}. IC~1396W: \cite{2003A&A...407..207F}.
IC~1396A: \cite{2004ApJS..154..385R,2009ApJ...690..683R}). More and more young stellar objects (YSOs) have been detected with CHANDRA, SPITZER and other bands \citep{2009AJ....138....7M,2011MNRAS.415..103B,2012AJ....143...61N}.

The pioneering InfraRed Astronomical Satellite (IRAS) all-sky survey covers more than $96\%$ of the entire sky in four photometric bands at 12, 25, 60 and 100 $\mu m$. The IRAS Sky Survey Atlas (ISSA) has shown that mid- and far-IR census is essential for studying the dust embedded objects in IC~1396. 
2MASS is the first near-IR survey that made uniformly calibrated
observations of the entire sky in the $J\ (1.25\ \mu m), H\ (1.65\ \mu
m)$, and $K_S$ $(2.16\ \mu m)$ bands with a pixel size of
2$^{\prime\prime}$.0. Sources brighter than about 1~mJy in each band
were detected with a signal-to-noise ratio greater than 10, which
leads to a photometric completeness to 15.9, 15.0, and 14.3 mag,
respectively, for each band in unconfused regions. For details about
the 2MASS survey and the 2MASS Point Source Catalog (PSC), please
consult the 2MASS Explanatory
Supplement\footnote[1]{http://www.ipac.caltech.edu/2mass/releases/allsky/doc/explsup.html}.

Near-IR large area surveys and 2MASS provide an opportunity to
investigate the sources with excessive emission, but their mid- and
far-IR counterparts are rare and are hard to be uniquely identified in
the IRAS catalogue, preventing an efficient search of objects
surrounded by dust. AKARI, a satellite operated by Japan, fulfills
the need for a new mid- and far-IR whole sky survey with better
sensitivity and higher spatial resolution. Since its launch on 2006
February 21, AKARI has mapped 96\% of the entire sky in mid and far-IR using two instruments on board: the InfraRed Camera
\citep[IRC;][]{2007PASJ...59S.401O} and Far-Infrared Surveyor
\citep[FIS;][]{2007PASJ...59S.389K}. The FIS swept about 94\% of the
whole sky more than twice at 65, 90, 140, and 160 $\mu m$ wavebands,
while IRC swept more than 90\% of the whole sky more than twice
using two filter bands centered at 9 $\mu m$ (\textsl{S9W}, 7 - 12
$\mu m$) and 18 $\mu m$ (\textsl{L18W}, 14 - 25 $\mu m$)
\citep{2010A&A...514A...1I}.

Most young stellar objects (YSOs) present an IR excess, which is interpreted in terms of the presence of a circumstellar disk. We
focus on the spatial distribution of the sources with excessive emission and try to elucidate the mode of star formation in IC~1396.
This paper is organized as follows. We present in Section~2 the retrieval of the archived 2MASS and AKARI
data, optical spectroscopy and its data reduction. In Section~3, we present how we explored the four subclusters spatially coincident
with the bright rims of IC~1396. Optical identification and SED classification of the sample sources follows in Section~4. The results
achieved are discussed in Section~5 and summarized in Section~6.

\section{Data Acquisition and Analysis}
\label{sect:Dat}

\subsection{Archival Data}

Archived data from the 2MASS PSC were used in this work. To guarantee
the reliability of the data, we employed the following strict
requirements in the sample selection, which are revised based on the
criteria presented by \cite{2005A&A...431..925L}: 1)The photometric uncertainties for each extracted 2MASS point source should be less than or equal to 0.1 mag at all three bands ($[JHK_S]_{-cmsig} \leqslant 0.1$). 2)~Only sources with a Ks-band signal-to-noise ratio above 15 are selected.

The sample of excessive emission sources selected based on the 2MASS data is cross-identified with the AKARI (ASTRO-F) IRC PSC\footnote[2]{http://darts.isas.jaxa.jp/astro/akari/cas.html} using
the simple positional correlation method. Only sources with valid
S9W data (fQual\_09 equals to 3) were considered. Based on the statistical result of the relationship between the positional offset and the numbers of counterparts both detected with the AKARI IRC PSC and 2MASS PSC, we found the chance of false matches becomes large when the tolerance radius is larger than $3^{\prime\prime}$. So a positional
tolerance of $3^{\prime\prime}$ is adopted. If more than one AKARI
source are found within the tolerance radius, only the closest one is adopted. The flux of the IRC
PSC sources were then converted to apparent magnitude with the
following equation:

$$m = -2.5 \times log_{10}(\displaystyle\frac{Flux}{Flux_0})$$
where $Flux_0$ (zero magnitude flux) is 56.262 $\pm$ 0.8214 Jy and
12.00 $\pm$ 0.1751 Jy for 9 $\mu m$ and 18~$\mu m$ bands,
respectively.

\subsection{Optical Spectroscopy}

Spectroscopic observations of all the sample sources in IC~1396 with USNO R magnitudes brighter than 17.0 were undertaken on the 2.16 m
optical telescope of the National Astronomical Observatories of the Chinese Academy of Sciences (NAOC).
The OMR (Optomechanics Research Inc.) and the detector PI 1340 $\times$ 400 CCD were used in both runs of observations. The
low-resolution spectroscopy (with dispersion of 200 \AA~mm$^{-1}$, 4.8 \AA~ pixel$^{-1}$, and 2$^{\prime\prime}$.5 slit)
 centered at 6300~\AA~ has been carried out on September 12, December 11, 2009, and October
3 to 4, 2010.

The spectral data were reduced following standard procedures in the NOAO Image Reduction and Analysis Facility (IRAF, version 2.11)
software packages. The CCD reduction includes bias and flat-field
correction, nebular and sky background subtraction, and cosmic ray
removal. Wavelength calibration was performed based on Helium-Argon
lamps exposed at both the beginning and the end of the observations
each night. Flux calibration of each spectrum was conducted based on observations of at least two of the KPNO spectral standards
\citep{1988ApJ...328..315M} each night. The atmospheric extinction is corrected by the mean extinction coefficients measured for the
Xing-Long Station, where the 2.16m telescope is located, by the Beijing-Arizona-Taiwan-Connecticut (BATC) multicolor survey.

\begin{figure}[hb]
   \centering
   \includegraphics[width=7.5cm, angle=0]{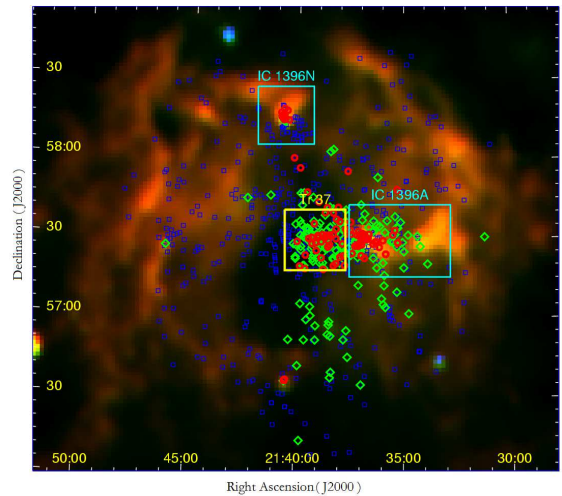}

   \begin{minipage}[]{85mm}

   \caption{Distribution of known YSOs in IC~1396. Historical studies were limited to the densest parts: IC~1396A, IC~1396N and Tr~37. Red circles indicate weak--line T Tauri stars and CLASS II/III sources, while green diamonds indicate classical T Tauri stars and CLASS I sources. The background image is a trichromatic image generated from IRAS 25 (blue), 60 (green), and 100 (red) $\mu m$ maps.}
   \label{hist}

    \end{minipage}

   \end{figure}

\section{Clustering of star formation in IC~1396}

Historical studies revealed the existence of many YSO candidates and most of them are found to be congregated to the densest parts of IC~1396. Figure \ref{hist} shows the distribution of YSOs in the literature as mentioned in Section~1. Its border shows the region studied in this paper. It extends from $ 21^{h} 28^{m} $ to $21^{h} 52^{m} $ in right ascension and from $56^\circ$ to $59^\circ$ in declination, centered on R.A. = $ 21^{h} 40^{m} 00^{s}, {\rm Dec} = 57^\circ 30^{\prime}00^{\prime\prime}$ (J2000.0), which is
believed to encompass all the compact subclusters to their full
extent.

\subsection{Color-color Diagram}

The 2MASS database contains more than half a million
photometric detections in this region. We narrowed down the
catalogue to 108,966 sources using our select criteria mentioned above.

\begin{figure}[htb]
   \centering
   \includegraphics[width=9.5cm, angle=0]{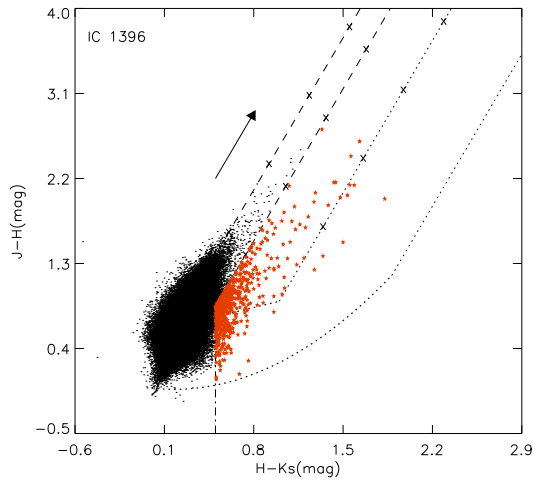}

   \begin{minipage}[]{85mm}
  \caption{$(J-H)$ vs. $(H-K_S)$ C-C diagram of IC~1396. The sample sources are denoted as dots.
Selected IR excessive emission sources are labeled in red. Black
solid lines
 represent the loci of the MS dwarfs and giant stars \citep{1988PASP..100.1134B}. Two paralleled dashed lines define the reddening band for normal stars. The arrow
 shows a reddening vector of $A_v=5$ mag \citep{1985ApJ...288..618R}. The left dotted line indicates the locus
 of dereddened T~Tauri star \citep{1997AJ....114..288M} and its reddening band boundary. The right dotted line indicates the locus of
 dereddened HAeBe \citep{1992ApJ...393..278L} and its reddening direction. Crosses were over plotted with an interval corresponding to 5 mag
 of visual extinction.\label{CCD1}} \end{minipage}
   \end{figure}

2MASS $JHK_S$ color-color (C-C) diagram is widely used to select sources with IR excessive emissions. Figure \ref{CCD1} shows the C-C diagram for IC~1396. All the 2MASS sources that match our criteria are put into the $JHK_S$ C-C diagram and denoted as dots. We define 598 objects as YSO candidates whose colors place them below the right line of the normal star reddening band and $H-K_S > 0.5$. Those sources are selected for they possess intrinsic color excesses likely originating from emission of circumstellar dust, commensurate with their embedded nature. The color distinction helps to eliminate foreground field stars and narrowed the YSO candidates sample.

   \begin{figure}[htb]
   \centering
   \includegraphics[width=120mm]{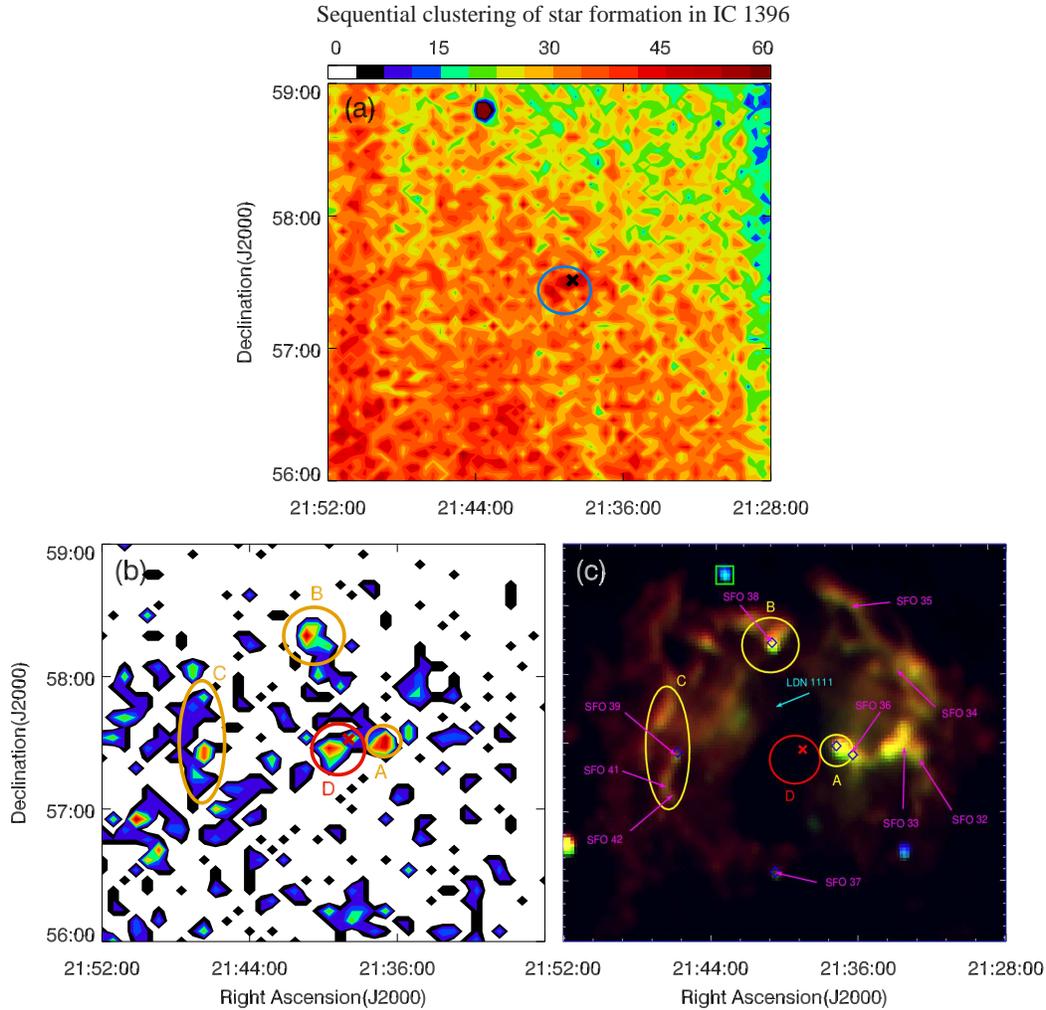}

\caption{(a) Surface density distribution of the 2MASS sources toward IC~1396. The densest region at the center marked with circle is consistent with the position of the open cluster Tr~37, and the cross indicates HD~206267
(R.A.= $21^h38^m57^s$, decl.= $+57^\circ 29^{\prime}20^{\prime\prime}(J2000)$. The hole on the top left corner is formed for a saturated bright source, which is marked with green box in panel (c). The color bar indicates the number of sources in every 0.0025 square degrees. (b) Spatial distribution for 598 2MASS excessive emission sources. Red cross indicates the exciting star HD~206267. Yellow circles mark the four densest regions of YSO candidates, which harbor active star formation. (c) Trichromatic image of IC~1396 generated from IRAS 25 (blue), 60 (green), and 100 (red) $\mu m$ maps. Bright rims \citep{1991ApJS...77...59S} and BOLOCAM data (blue diamonds) are plotted. Red cross indicates the main exciting source HD~206267.}

\label{DMA}
\end{figure}

\subsection{Surface Density Distribution}

The spatial distribution of all the 2MASS sources and YSO candidates are presented in Figure \ref{DMA}. On the scale of the entire IC~1396, significant stellar density
enhancement in the center region is easily distinguished in the
stellar surface density distribution of all the 2MASS sources~(in Figure~\ref{DMA} (a)). There are about 59 sources in 0.0025 deg$^2$ at the densest region, which coincides with the young open cluster Tr~37. But we can't find the structure of the H~\textsc{ii} region in this panel.

In panel (b), the spatial distribution of the YSO candidates, there are four densest regions revealed. Compared with the 100~$\mu m$ image from IRAS, the A, B, C subclusters correspond to the bright rims \citep{1991ApJS...77...59S} marked in Figure \ref{DMA} (c), while the central subcluster D is at the location of open cluster Tr~37. Otherwise, region A and region B correspond to IC~1396~A and IC~1396~N, centered at 21$^h$36$^m$54$^s$,
57$^\circ33^{\prime}00^{\prime\prime}$ (J2000.0) and
21$^h$40$^m$42$^s$, 58$^\circ15^{\prime}14^{\prime\prime}$
(J2000.0), respectively. Region C is the most scattered one, and
contains several stellar aggregates. Among them the most compact
and biggest one is centered at 21$^h$46$^m$15$^s$,
57$^\circ25^{\prime}00^{\prime\prime}$ (J2000.0).
Compared to the outer regions A, B, and C, region D is hardly to be found in ISSA image, because it has already lost dust protection in IR bands for the radiation of the exciting star HD~206267. But it still contains a number of YSO candidates.

\section{Identification and classification of excessive emission sources in IC~1396}
To further investigate the scenario of star formation in IC~1396, AKARI IRC PSC was employed. Mid-IR data are more sensitive to cold dusts in the circumstellar disks and are essential for SED fitting. All the excessive emission sources selected based on the 2MASS C-C diagram were cross-identified with the AKARI IRC PSC, which resulted in a sample of 44 sources. Except 4 classical Be stars (CBe) and 2 Carbon stars identified with SIMBAD, all of
the sample sources are presented in Table \ref{ALL}, which presents the sequences number of these sources, 2MASS PSC coordinates (J2000),
USNO V magnitudes, EW[H$\alpha$], EW[Li], spectral types, and the classification based on SEDs.

  \begin{figure}[!b]
   \centering
   \includegraphics[width=110mm, angle=0]{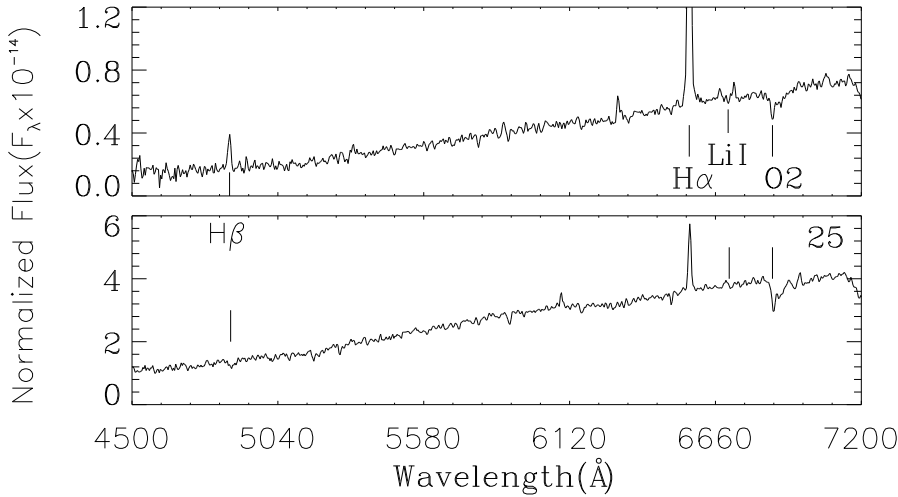}
   \caption{Typical WTTS (No.21) and CTTS (No.25) spectra in IC~1396, showing the Balmer series in emission (mainly H$\alpha$) and Li absorption at 6707 \AA\ .}
   \label{SPECTRA}
   \end{figure}

\subsection{Optical identification}

To identify and classify the sample sources, optical spectra were obtained with the 2.16 m optical telescope of the NAOC. Combined with equivalent width measurements of H$\alpha$ (EW[H$\alpha$]), all the sources with spectral types earlier than F5 and H$\alpha$ in emission were identified as Herbig Ae/Be stars (HAeBe).  Classical T Tauri stars (CTTS) are, on the other hand, selected based
on the criteria that a CTTS shows EW[H$\alpha$] value greater than 3 \AA\ for K0 -- K5 stars, EW[H$\alpha$] value greater than
10 \AA\ for K7 -- M2.5 stars, EW[H$\alpha$] value greater than 20 \AA\ for M3
-- M5.5 stars, and EW[H$\alpha$] value greater than 40 \AA\ for M6 -- M7.5 stars
\citep{2003ApJ...582.1109W}. However, stars with values of EW[H$\alpha$] below these levels are not necessarily weak-line T Tauri stars (WTTS) as they share very similar SEDs with MS stars. Further identification based on the detection of Li I 6707 \AA~absorption is necessary \citep{1998AJ....115..351M}. Except 4 CBe, 23 stars out of the 33 observed sources are identified as YSOs due to the presence of prominent H$\alpha$ emission. Among these, there are 5 CTTS, 7 HAeBe, and 8 WTTS. For the left 3, no prominent Li I absorption is detected because of low Signal to Noise Ratio of the optical spectra. However, the existence of strong
H$\alpha$ emission and excessive emission in IR seems to corroborate their nature as YSOs. Figure \ref{SPECTRA} illustrates the sample spectra of WTTS and CTTS, respectively. Among the sample of 33 observed sources, 27 were found to be strong H$\alpha$ emission stars. And 3 HAeBe, 1 CTTS and 13 WTTS are newly discovered in IC~1396.

\subsection{Classification based on SED fitting}

We tried to determine the evolutionary status of the sample sources based on their SEDs. A grid of 200,000 YSO models was developed \citep{2006ApJS..167..256R}, spanning a wide range of evolutionary stages for different stellar masses, to model the SED from optical to millimeter wavelengths. This archive transfer provides a linear regression tool which can select all model SEDs that fit the observed SED better than a specified $\chi^2$ \citep{2006ApJS..167..256R}.

On the basis of the ``four staged" star formation scenario proposed
by \cite{shu87}, \cite{1987IAUS..115....1L} developed a widely used
classification scheme for YSOs, primarily based on their SEDs. With
an evolutionary sequence from early type to late type, YSOs were
classified into Class I to III. \cite{2006ApJS..167..256R} presented
a classification scheme that is essentially analogous to the class
scheme, but refers to the actual evolutionary stage of the object
based on physical properties like disk mass and envelope accretion
rate. However, in view of the differences between observable and
physical properties, ages fitted by the tool and the slope of its
near/mid-IR SED are our primary reference standard. Class I refers
to those objects that have $Age\ \approx10^5$ yr and
$Slope_{near/mid-IR}> $ 0, Class II refers to $Age\ \approx10^6$ yr
and $Slope_{near/mid-IR}\le $ 0, and Class III is for objects whose
SED is similar to a black-body spectrum.

  \begin{figure}[ht]
   \centering
   \includegraphics[width=110mm, angle=0]{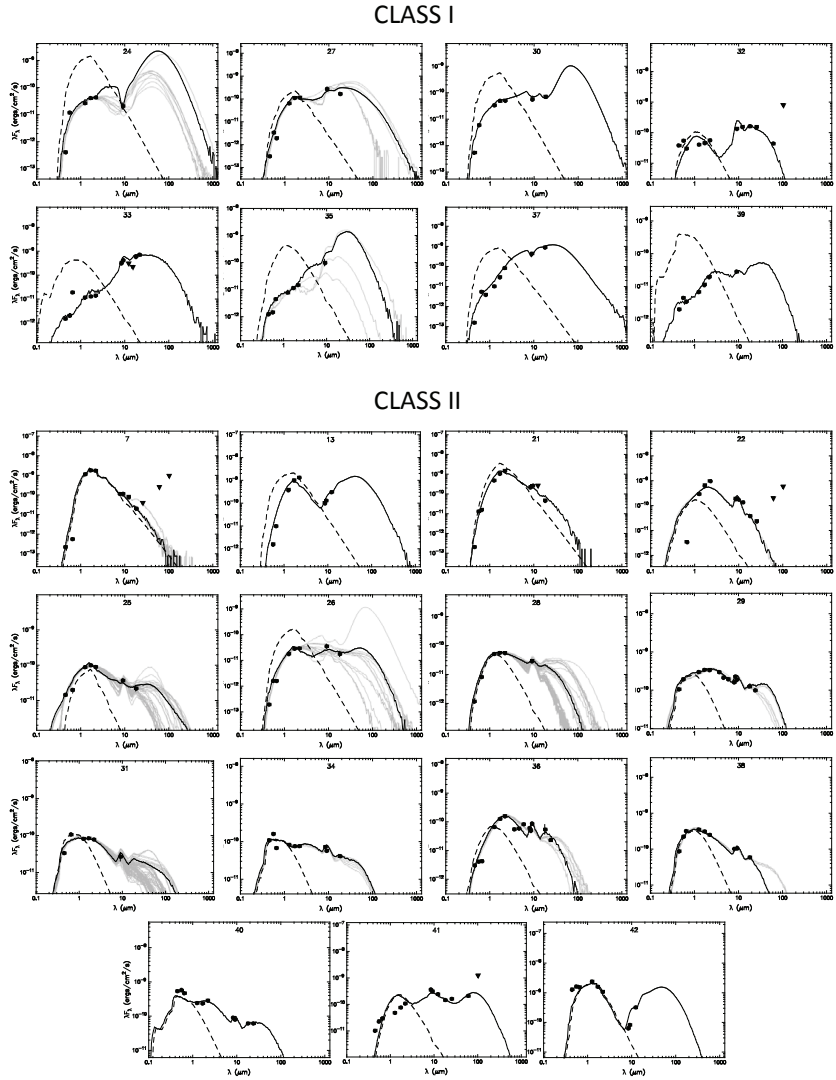}
   \caption{Results of the SED fitting for the sample sources. The solid black curve gives the best model fitting of the data points with the smallest $\chi^2$. The gray curves represent the other potential model fitting results with $\chi^2$/N-$\chi^2_{best}/N$$<3$. The dashed curve indicates the photosphere that is used as input for the radiative transfer code. The black dots show the measured fluxes currently available. The filled upside-down triangles indicate upper limit detection in each of the corresponding bands.}
   \label{SED}
   \end{figure}

Multi-wavelength online data are used for SED fitting. Besides 2MASS PSC and AKARI IRC/FIS data, we used: 1) BVR photometry from the Naval Observatory Merged Astrometric Dataset (NOMAD); 2) the mid-IR A (8.28 $\mu m$), C (12.13 $\mu m$) and D (14.65 $\mu m$) provided by the MSX6C IR PSC; 3) the far-IR IRAS Point Sources photometry at 12, 25, 60 and 100$\mu m$; 4) IRAC and MIPS photometry at 3.6, 4.5, 5.8, 8 and 24m for a handful of sources. As a result, we end up SED fitting with a photometric catalogue with large wavelength coverage. Though several sources lack data at far-IR bands, it is still crucial to investigate the PMS nature of sample sources in terms of IR excess in their SEDs.

The evolutionary status of all the YSO candidates listed in Table \ref{ALL} further confirmed by results of the SED fitting. Figure \ref{SED} illustrates the SED of the CLASS I and CLASS II sources of the sample. Based on the SED fitting, all the sample sources indicate masses of $<$ 5 M$_{\odot}$ and ages of $<$ 3 Myr, which is in good accordance with the results of \cite{2012AJ....143...61N} and \cite{2010ApJ...717.1067C}.

\clearpage

   \begin{longtable}{cccccccccc}
\multicolumn{10}{c}{\tablename\ \thetable{}: Optical identification and classification of the sample sources toward IC~1396.
\label{ALL} } \\
\hline
\hline
      ID & R.A.& Dec. &      V & \multicolumn{2}{c}{EW} & Sp. & \multicolumn{3}{c}{Class}  \\
   \cline{2-3}\cline{5-6}\cline{8-10}
      &\multicolumn{2}{c}{J2000}&[mag]&$[H\alpha]$&[Li]&&Spectra&C-C diagrams&SEDs \\\hline

 \endfirsthead
\hline\hline

      ID & R.A.& Dec. &      V & \multicolumn{2}{c}{EW} & Sp. & \multicolumn{3}{c}{Class}  \\
   \cline{2-3}\cline{5-6}\cline{8-10}
      &\multicolumn{2}{c}{J2000}&[mag]&$[H\alpha]$&[Li]&&Spectra&C-C diagrams&SEDs \\\hline
\endhead
\hline
\endfoot

\endlastfoot

         1$^\ast$ &   21 42 33.57 &    58 14 39.11 &      12.21 &      -12.1 &     ...    &         B2 &        CBe &       CBe  &    ...     \\
         2$^\ast$ &   21 48 21.16 &    56 09 20.15 &      12.41 &      -9.05 &      ...   &         B4 &        CBe &       CBe  &       ...  \\
         3$^\ast$ &   21 36 59.63 &    58 08 24.72 &      12.21 &      -27.9 &      ...   &         B3 &        CBe &       CBe  &       ...  \\
         4 &   21 42 24.16 &    57 44 09.96 &      14.77 &      -4.99 &      ...   &         O9 &        CBe &       CBe  &       ...  \\
         5 &   21 29 23.95 &    56 19 49.43 &      14.73 &      -2.15 &       0.57 &         M4 &       WTTS &      WTTS  &        III \\
         6 &   21 47 15.45 &    56 23 45.95 &      15.63 &      -6.18 &       0.08 &         M4 &       WTTS &      WTTS  &        III \\
         7 &   21 43 33.09 &    57 25 25.31 &      24.61 &      ...   &      ...   &     ...    &    ...     &      WTTS  &         II \\
         8 &   21 46 35.97 &    57 49 31.79 &      18.07 &      ...   &      ...   &      ...   &      ...   &      WTTS  &        III \\
         9$^\ast$ &   21 43 08.73 &    57 11 58.20 &       16.1 &      -20.5 &       2.02 &         M7 &       WTTS &      WTTS  &        III \\
        10$^\ast$ &   21 34 15.59 &    56 18 11.88 &       16.1 &      -1.97 &       0.35 &         M6 &       WTTS &      WTTS  &        III \\
        11$^\ast$ &   21 40 03.86 &    57 42 18.71 &      17.41 &       ...  &       ...  &         M3 &        ... &      WTTS  &        III \\
        12$^\ast$ &   21 45 34.12 &    56 23 42.72 &      16.81 &       ...  &       ...  &         M2 &       ...  &      WTTS  &        III \\
        13$^\ast$ &   21 32 44.54 &    56 21 25.56 &       16.1 &      -4.56 &       0.86 &         M7 &       WTTS &      WTTS  &         II \\
        14 &   21 46 41.23 &    57 00 10.80 &      14.97 &      -3.38 &       0.48 &         M5 &       WTTS &      WTTS  &        III \\
        15$^\ast$ &   21 39 10.63 &    57 06 47.16 &      16.01 &      -2.36 &       0.53 &         M4 &       WTTS &      WTTS  &        III \\
        16$^\ast$ &   21 50 21.50 &    56 53 10.32 &      20.61 &       ...  &       ...  &       ...  &       ... &      WTTS  &        III \\
        17 &   21 43 16.43 &    56 01 42.60 &       14.3 &      -10.2 &        ... &         M9 &       WTTS &      WTTS  &        III \\
        18 &   21 42 47.61 &    58 57 20.52 &      13.11 &        ... &       ...  &         M8 &       ...  &      WTTS  &        III \\
        19 &   21 42 12.04 &    58 25 12.00 &      26.41 &        ... &       ...  &         M8 &       ...  &      WTTS  &        III \\
        20$^\ast$ &   21 39 06.38 &    57 43 51.60 &      16.95 &      -14.1 &       ...  &         M7 &       WTTS &      WTTS  &        III \\
        21 &   21 38 42.31 &    57 30 27.71 &      16.41 &      -12.9 &       ...  &         M4 &       WTTS &      WTTS  &         II \\
        22 &   21 39 09.33 &    58 38 53.87 &      16.41 &       ...  &       ...  &         M7 &      ...   &      WTTS  &         II \\
        23$^\ast$ &   21 43 51.47 &    58 15 07.92 &      19.73 &       ...  &       ...  &       ...  &       ...  &      WTTS  &        III \\
        24$^\ast$ &   21 48 11.75 &    57 59 41.64 &       15.7 &      -86.3 &       0.84 &         F8 &       CTTS &       CTTS &          I \\
        25 &   21 36 49.56 &    57 48 23.39 &      14.85 &      -6.52 &       0.02 &         G6 &       CTTS &       CTTS &         II \\
        26 &   21 45 05.87 &    57 11 38.75 &      18.21 &        ... &        ... &        ... &       ...  &       CTTS &         II \\
        27 &   21 34 19.63 &    57 30 02.52 &       15.7 &      -45.1 &       0.23 &         F7 &       CTTS &       CTTS &          I \\
        28 &   21 35 43.60 &    57 03 47.52 &      14.81 &      -2.92 &       0.23 &         K3 &       WTTS &       CTTS &         II \\
        29$^\ast$\footnotemark[1] &   21 38 17.32 &    57 31 22.07 &         13 &      -14.5 &       0.11 &         F9 &       CTTS &       CTTS &         II \\
        30 &   21 46 00.26 &    57 23 09.60 &      18.35 &       ...  &        ... &        ... &        ... &       CTTS &          I \\
        31 &   21 45 54.07 &    57 28 18.48 &       14.4 &      -9.65 &        0.1 &         K4 &       CTTS &       CTTS &         II \\
        32 &   21 30 22.84 &    58 28 51.95 &      14.37 &      -40.5 &       0.31 &         A9 &      HAeBe &     HAeBe  &          I \\
        33 &   21 49 38.27 &    56 54 36.72 &       15.7 &       -141 &       0.12 &         F2 &      HAeBe &     HAeBe  &          I \\
        34 &   21 35 19.15 &    57 36 38.15 &      15.21 &      -6.98 &       0.01 &         A8 &      HAeBe &     HAeBe  &         II \\
        35$^\ast$ &   21 29 58.03 &    56 28 50.51 &       17.5 &       ...  &       ...  &        ... &       ...  &     HAeBe  &          I \\
        36\footnotemark[2] &   21 36 39.14 &    57 29 53.16 &         17 &        ... &       ...  &        ... &       ... &     HAeBe  &         II \\
        37 &   21 46 07.12 &    57 26 31.91 &      17.21 &       ...  &       ...  &        ... &       ...  &     HAeBe  &          I \\
        38 &   21 45 02.32 &    56 49 51.60 &      14.21 &       2.35 &       0.25 &         F9 &       ...  &     HAeBe  &         II \\
        39 &   21 45 24.55 &    57 55 49.08 &       16.1 &      -11.1 &       0.79 &         F3 &      HAeBe &     HAeBe  &          I \\
        40 &   21 38 08.44 &    57 26 47.75 &      14.15 &         -5 &        0.1 &         A2 &      HAeBe &     HAeBe  &         II \\
        41 &   21 33 17.78 &    57 48 13.31 &      14.53 &      -62.6 &       0.13 &         F3 &      HAeBe &     HAeBe  &         II \\
        42 &   21 51 00.57 &    56 21 19.44 &      10.24 &      -43.1 &       ...  &         F3 &      HAeBe &     HAeBe  &         II \\
        43 &   21 43 56.92 &    58 35 45.95 &      18.17 &        ... &       ...  &       ...  &      ...   &      Carb. &        ... \\
        44 &   21 38 28.34 &    57 08 19.32 &      16.91 &       ...  &        ... &         M7 &      ...   &      Carb. &        ... \\\hline
        \multicolumn{10}{l}{\footnotesize{$^\ast$ Sources not detected with IRC 18 $\mu m$}}\\
        \multicolumn{10}{l}{\footnotesize{$^1$ IC~1396 A:$\theta$ -- identified as CTTS from \cite{2006AJ....132.2135S}.}}\\
        \multicolumn{10}{l}{\footnotesize{$^2$ Identified as CLASS II source from \cite{2009ApJ...690..683R}.}}
\end{longtable}
\clearpage
\section{Discussion}

\cite{1990AJ.....99.1536M} suggested that stars of the first generation in IC~1396 reached the main sequence at 7 Myr ago, and \cite{2005AJ....130..188S} gave an age of 4 Myr for the young open cluster Tr~37 based on optical photometry and theoretical isochrones. The expansion of the H~\textsc{ii} region has created a swept-up
shell and resulted in a compressed molecular ring around its
periphery \citep{2012AJ....143...61N}. Based on the physical scale of the ring, \cite{1995ApJ...447..721P} suggested
a dynamical age of 2 -- 3 Myr. A number of H$\alpha$ stars with ages younger than 3 Myr were detected by \cite{2012AJ....143...61N}. They are spatially associated with the bright rims. Based on the disparity of ages for various groups of stars, we tentatively suggest a scenario of sequential or episodic star formation in IC~1396.

\begin{figure}[htb]
   \centering
   \includegraphics[width=9.0cm, angle=0]{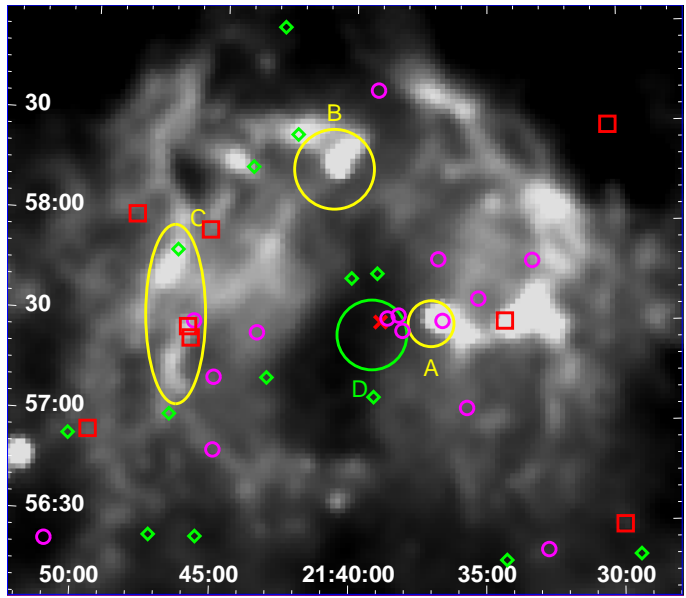}

   \begin{minipage}[]{85mm}

   \caption{Distribution of the sample sources classified with SEDs. CLASS I, II, and III sources are indicated with red boxes, magenta circles, and green diamonds, respectively. The red cross represents the exciting star HD~206267. The background image is the IRAS 100$\mu m$ map of IC~1396}
   \label{SEDD} \end{minipage}
   \end{figure}

In this work, we provide new additional evidence to support a scenario of sequential or episodic formation of stellar clusters in IC~1396. Figure 6 presents the IRAS 100 $\mu$m mosaic of IC~1396, on which the sample sources classified based on SEDs are overlaid with different symbols. Although the number of Class I sources identified in this study is small, all the CLASS I sources are found to be located in the bright rims or the molecular ring of IC~1396, while YSOs at later evolutionary stages (e.g. CLASS III) are congregated to the central part of the H~\textsc{ii} region. Furthermore, all the identified CLASS I and CLASS II sources are located in the surrounding sub-clusters rather than the central open cluster Tr~37. All these are consistent with an episodical nature of star and cluster formation in this H~\textsc{ii} region. A direct consequence of episodic star formation is that different generations of stars occupy distinct territories.

\begin{figure}[h]
   \centering
   \includegraphics[width=9.0cm, angle=0]{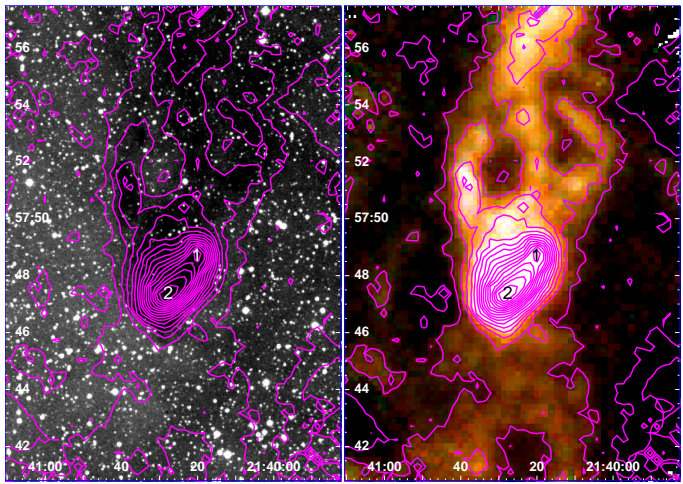}

   \begin{minipage}[]{85mm}

   \caption{Left: DSS 2 blue band image of LDN 1111. Right: Color composite image of the dark cloud LDN 1111 in 250 $\mu m$ (blue), 350 $\mu m$ (green) and 500 $\mu m$ (red). Magenta contours in both two images are generated from the Herschel SPIRE image in 500 $\mu m$ band.}
   \label{L1111}
   \end{minipage}

   \end{figure}

A spatial gap seems to exist between the central open cluster and the surrounding subclusters in both the optical and the near-IR, which is believed to be the results of discontinuous star formation in this region. To further investigate the interstellar materials within the gap, Herschel SPIRE data were employed. As shown in Figure~\ref{DMA} (c), there is an optically dark cloud between Tr 37 and subcluster B. This dark cloud was catalogued by \cite{1962ApJS....7....1L} as LND~1111, which is highly extinct both in the optical and the near-IR. The right panel of Figure~\ref{L1111} presents a composite image of LND~1111, which was compiled with the SPIRE 250 $\mu m$ (blue), 350 $\mu m$ (green), and 500 $\mu m$ (red) imaging data. It is evident that the dark globule corresponding to LND~1111 in the DSS-2 blue band image (left panel in Figure~\ref{L1111}) is bright in emission in the far-IR. This indicates the existence of large amount cold dusts. The overlaid contours generated based on the Herschel SPIRE image indicate the presence of two far-IR cores, which are likely nurseries of new generation stars. Therefore, we infer that new generations of star formation is going on in IC~1396, and star formation in this region is not continuous but sequential or episodic. Further investigations are necessary to explore if new generations of star and cluster formation in IC~1396, both in the vacant bubble and the working interface of the H$\textsc{ii}$ region, has a triggered origin or not.

\section{Summary}

We present in this paper a comprehensive study of the H~\textsc{ii} region IC~1396 and its star formation activity.
Excessive emission sources are selected based on the archived 2MASS data and their cross-identification with AKARI IRC PSC. SED fittings are employed to classify the IR sources. In the target region, 8 CLASS I, 15 CLASS II and 15 CLASS III sources were classified. Optical identification of the sample sources with USNO R magnitudes brighter than 17.0  corroborates with the results from the classification of the IR sources based on SED fitting. Among the sub-sample of 33 observed sources, 27 were found to present strong H$\alpha$ emission. However, 3 HAeBe, 1 CTTS and 13 WTTS are newly discovered.

The spatial distribution of the IR excessive emission sources selected based on their 2MASS colors reveals four sub-clusters toward IC 1396. One is spatially in association
with the open cluster Trumpler 37 that hosts the exciting source of the H~\textsc{ii} region. The other three are found to be spatially coincident with the bright rims the surroundings.
All the identified CLASS I and CLASS II sources are found to be located in the surrounding sub-clusters rather than the central young open cluster Tr~37, which is primarily a
congregation of CLASS III sources. The surrounding sub-clusters are thus believed to much younger due to the spatial distribution of IR sources at different evolutionary stages. This is consistent with a scenario of sequential star formation in this region. Imaging data of a dark patch in IC~1396 by Herschel SPIRE indicate the presence of two far-IR cores, which are likely evidence for new generations of star formation. Therefore, we infer that star formation in IC~1396 was not continuous but sequential and/or episodic.

\normalem
\begin{acknowledgements}
We appreciate very much the helpful comments and suggestions from the referee.
This work employed data from the 2MASS, AKARI and other
database. Our investigation is supported by funding from the
National Natural Science Foundation of China through grant NSFC
11073027, 11003021 and the Department of International Cooperation of the
Ministry of Science and Technology of China through grant
2010DFA02710.
\end{acknowledgements}

\label{lastpage}

\end{document}